\documentclass{osa-article}

\journal{osajournal}

\articletype{Research Article}

\begin{document}

\title{Improving the sensitivity of planar Fabry-P\'erot cavities via adaptive optics and mode filtering} %

\author{Jakub Czuchnowski,\authormark{1,\S} Robert Prevedel,\authormark{1-4,*}}

\address{\authormark{1}Cell Biology and Biophysics Unit, European Molecular Biology Laboratory, Heidelberg, Germany\\
\authormark{2}Developmental Biology Unit, European Molecular Biology Laboratory, Heidelberg, Germany\\
\authormark{3}Epigenetics and Neurobiology Unit, European Molecular Biology Laboratory, Monterotondo, Italy\\
\authormark{4}Molecular Medicine Partnership Unit, European Molecular Biology Laboratory, Heidelberg, Germany
\authormark{\S}Collaboration for joint PhD degree between EMBL and Heidelberg University, Faculty of Biosciences, Germany}
\email{\authormark{*}prevedel@embl.de}

\begin{abstract}
Fabry-P\'erot (FP) cavities are fundamental and ubiquitous optical elements frequently used in various sensing applications. Here, we introduce a general theoretical framework to study arbitrary light-cavity mode interactions for planar FPs and show how optical aberrations, intrinsic to the interrogating beam or due to imperfect cavities, reduce optical sensitivity by exciting higher order spatial modes in the cavity. We find that particular Zernike aberrations play a dominant role in sensitivity degradation, and that the general loss of sensitivity can be significantly recovered by appropriate wavefront correction or mode filtering. We then demonstrate our theoretical findings also experimentally and show that in practice the sensitivity of realistic planar FP sensors can be improved up to three-fold by a synergistic combination of adaptive optics and passive mode filtering.
\end{abstract}

\section{Introduction}
Fabry-P\'erot (FP) cavities are omnipresent in many important devices used in optics, such as in laser cavities or as narrowband wavelength filters in spectrometers. Their potential for ultra-high sensitivity measurements was recently demonstrated by LIGO with the detection of gravitational waves \cite{Abbott:16}. While most FP cavities come in a confocal configuration which makes them in general more stable, some recently emerging optical systems require \textit{planar} FP configurations. Among them are FP based pressure sensors employed in e.g. high-sensitivity photo-acoustic imaging \cite{Zhang:08}, or so-called Virtually Imaged Phased Arrays (VIPAs) which enable high-resolution spectroscopy applications such as Brilluoin microscopy \cite{Prevedel:19,Scarcelli:08}. As all-optical devices, the performance of FP cavities depends not only on FP intrinsic parameters such as mirror reflectivities, surface homogeneity, etc., but also on extrinsic properties such as the incident light beams that are used to interrogate them. Here in particular, wavefront distortion of the incident light beam due to imperfect optics and/or their alignment, or due to air turbulence have to be considered.

The effects of such wavefront distortions on FP sensitivity were previously studied for \textit{confocal} FP cavities, inside which the electro-magnetic field is quasi-stationary, i.e. not changing between individual reflections and round trips. In case of gravitational wave detectors, Bond \textit{et al.} \cite{Bond:11} have studied the role of mirror distortions, described using Zernike polynomials, and showed that they redistribute power among
Laguerre–Gaussian modes, an effect that was later experimentally demonstrated for several aberrations by Gatto \textit{et al.} \cite{Gatto:14}. In shorter cavities, Mah and Talghader \cite{Mah:19} as well as Takeno \textit{et al.} \cite{TAKENO20113197} have explored the use of FP cavities for aberration sensing by relying on spectral properties of the transmitted beam and intensity of the higher order spatial modes reflected away from the cavity, respectively, while Liu and Talghader \cite{LIU2006} examined the effects of imperfections in tunable micromirror cavities on Gaussian beams.

\begin{figure}
\includegraphics[width=11cm]{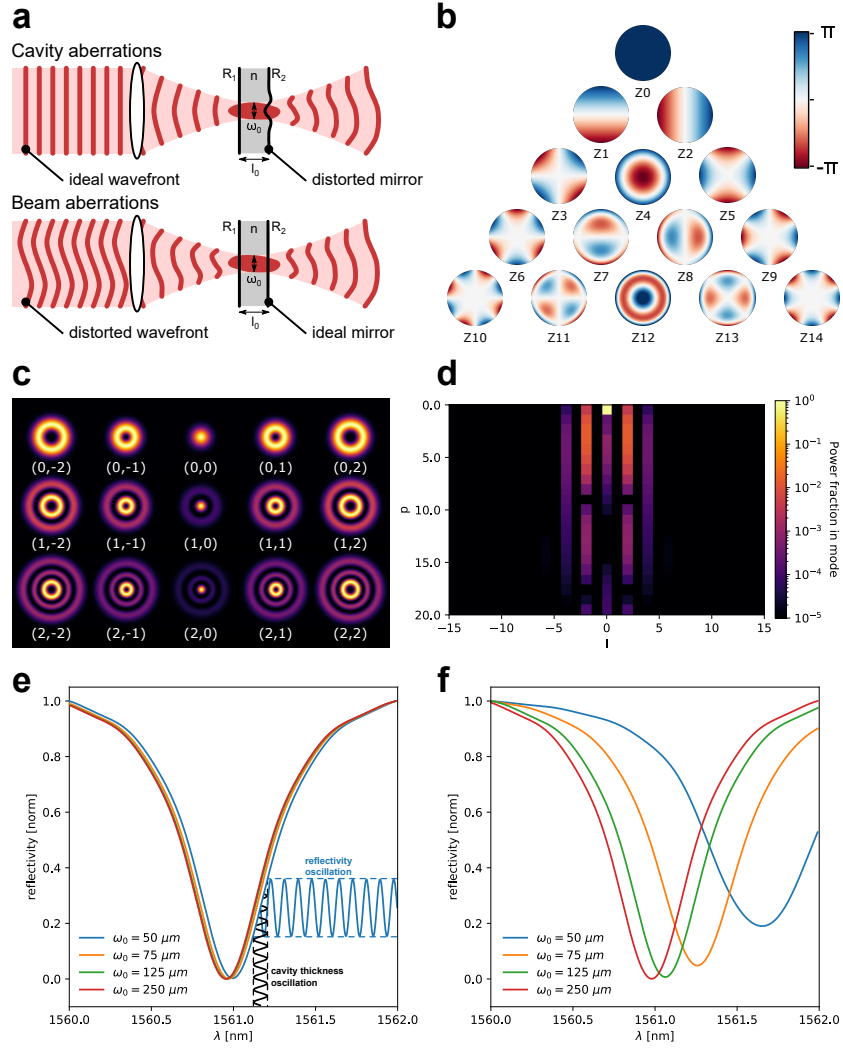}
\centering
\caption{\textbf{a.} A schematic conceptualising the effects of beam and cavity aberrations on the wavefront of the beam, where: $R_1,R_2$ - reflectivity of the two mirrors, $w_0$ - beam waist radius, $l_0$ - cavity thickness, n - refractive index of the elastic material inside the cavity.  \textbf{b.} Phase profile of the first 15 Zernike aberrations. \textbf{c.} Intensity profile of some of the low order $(l,p)$ Lauguerre-Gaussian modes. \textbf{d.} Decomposition of an arbitrarily aberrated beam into Laguerre-Gaussian modes. \textbf{e,f.} Effect of beam size on FPI transfer functions for two LG modes. \textbf{e} The fundamental $LG_{00}$ displays little dependence on increasing divergence of the beam (decreasing the spot size). Here we also show schematically how changes in cavity thickness are optically amplified at the bias wavelength. \textbf{f} In contrast, a higher order $LG_{55}$ mode displays pronounced TF distortions as a function of beam divergence. Additionally simulations for large spot sizes ($\omega_0 = 250 \ \mu m$) show little difference between the two modes confirming the calculations for an ideal FPI (\textbf{Appendix A}). For the simulations we choose the following cavity parameters: $l_0=20 \mu m$, $R_{1/2}=0.95$ which are commonly used design parameters for photoacoustic tomography systems
\cite{Zhang:08}.} 
\label{fig:AbbIntro}
\end{figure}

However, the effects of optical aberrations on \textit{planar} FP cavities, i.e. composed of flat mirrors, are generally much less studied, and in particular when interrogated by focused light (\textbf{Figure \ref{fig:AbbIntro}a}). Such a configuration has recently been shown to exhibit promising performance as highly sensitive pressure sensors for use in photo-acoustic imaging modalities \cite{Zhang:08,Jathoul:15}. Here, large efforts have been devoted to improving techniques for their manufacture \cite{Buchmann:17,Villringer:20}, however, a general theoretical framework to study the interaction of arbitrary light modes with their cavity counterparts is currently lacking. While recent work by Marques \textit{et al.} \cite{Marques:20} provides an accurate theoretical model for calculating reflectivity spectra for \textit{ideal} FP cavities illuminated with focused beams based on interference of plane waves, their model does not allow to draw general conclusions on the effects of aberrations on FP sensitivity.

In our work, we introduce an alternative general framework to study arbitrary light-cavity interactions that allows us to gain a broader understanding of the mechanisms by which optical aberrations degrade FP sensitivity. Our framework is based on extending the 'unfolded cavity approach' \cite{Varu:14,Abu-Safia:94} in order to account for beam aberrations by combining it with Gaussian Beam Mode Analysis (GBMA) \cite{Tsigaridas:03}. GBMA is based on Laguerre-Gaussian mode decomposition and beam propagation, which enables us to numerically investigate the coupling of arbitrarily aberrated Gaussian beams, expressed in terms of Zernike-modes, to Laguerre-Gaussian cavity modes (\textbf{Figure \ref{fig:AbbIntro}b,c}). This framework allows us to investigate how particular beam and cavity aberrations affect the FP's overall sensitivity. Our simulations show that the loss of sensitivity is generally caused by coupling and exciting higher-order cavity modes (\textbf{Figure \ref{fig:AbbIntro}d}), and that optimal sensitivity can be restored by optical mode filtering and active aberration correction techniques. We further demonstrate these effects experimentally and show that in practice the sensitivity of realistic planar FP sensors can be improved up to three-fold by a combination of adaptive optics and passive mode filtering.

\section{Fabry-P\'erot cavities for (photo-)acoustic sensing}
Since the main motivation of our work is connected to FP based pressure sensors and their use in photoacoustic imaging, we start by shortly introducing the concept and working principle of the technique. Photoacoustic tomography is a non-invasive deep-tissue imaging modality that uses light-induced acoustic waves to combine optical contrast with high-resolution ultrasound detection \cite{Wang:16}. To overcome limitations of classical ultrasound detectors, several optical methods for detection of photoacoustic waves were developed over the years (see \cite{Wissmeyer:18} for review). Here, the use of planar FP cavities has been particularly promising, as it combines high sensitivity with the ability to measure acoustic waves at well-defined spatial locations (given by the interrogating beam size on the sensor), which is important for high-resolution tomographic image reconstruction. In this approach an elastic FP cavity is formed by sandwiching a layer of elastomere (e.g. Parylene C) between two dichroic mirrors. The cavity can then deform elastically upon incidence of a pressure wave, thus modulating the position of the FP interferometer's transfer function (ITF) which depends on the (optical) thickness of the cavity. By tuning the interrogation laser wavelength to the point of maximum slope on the ITF (so-called bias wavelength) one obtains maximum sensor sensitivity, i.e. the incident acoustic wave is maximally amplified optically (\textbf{Figure \ref{fig:AbbIntro}e}). Experimentally, this approach has enabled acoustic sensing in the range of $100-10^6$ Pa with a broadband frequency response (bandwidths up to $\sim$40MHz) \cite{Zhang:08,Jathoul:15} .

\section{Theoretical approach}
We now proceed with introducing the main concepts to theoretically describe and analyze the dependence of sensitivity on optical and cavity aberrations. Since in theory the ITF is symmetrical, photoacoustic measurements can be made both on the falling as well as on the rising edge of the transfer function. We thus define two types of normalised optical sensitivity: $S_o^{+}$ and $S_o^{-}$ are the rising and falling edge sensitivity respectively,

\begin{equation}
S_o^{\pm} = \frac{\pm \frac{d}{d\lambda}I_{FPI}(\lambda)\bigg\rvert _{\lambda=\lambda_{opt}^{\pm}}}{I_{FPI}(\lambda_{opt}^{\pm})}
\label{eq:SO}
\end{equation}

where,

\begin{equation}
\lambda_{opt}^{\pm} = argmax \{\pm \frac{d}{d\lambda}I_{FPI}(\lambda)  \}
\end{equation}

is the bias wavelength, and

\begin{equation}
I_{FPI} (\lambda) = \iint \limits_{A} E_{FPI}(r,\phi,\lambda)^*E_{FPI}(r,\phi,\lambda)  \,dA 
\label{eq:I_FPI}
\end{equation}

is the Fabry-P\'erot interferometer transfer function, where A is the aperture over which the field is measured and $E_{FPI}$ is the interfering electric field given by:

\begin{equation}
E_{FPI}(r,\phi,\lambda) ={r_1^L}E(r,\phi,0,\lambda) + \sum \limits_{k=1}^{\infty}\beta_k E(r,\phi,z_k,\lambda'),
\label{eq:Efpi}
\end{equation}

with $z_k=2l_0k$, $\beta_k={(t_1)}^2{(r_1^R)}^{k-1}{(r_2^L)}^{k}$, $E(r,\phi,z,\lambda)$  denoting the electric field of the beam propagating in the cavity, $l_0$ the cavity length, $\lambda'$ the effective wavelength inside the cavity, $t_1$  the amplitude transmission coefficient for the first mirror, and $r_1^{R/L}$ as well as $r_2^L$ the amplitude reflection coefficients for the first mirror on the right/left side and second mirror for the left side, respectively.

This definition of $S_o$ allows normalisation for both laser relative intensity noise, as well as shot noise which are the dominant sources of noise in typical, realistic FPI systems and therefore is directly proportional to the SNR. Estimation of $S_o$ requires calculation of $E(r,\phi,z,\lambda)$ for different $z$ planes (\textbf{Equation \ref{eq:Efpi}}). This cannot be done analytically for aberrated Gaussian beams in general and thus requires a new theoretical framework. A number of approaches exist including numerically solving the Fresnel integral\cite{Sears:49}, or the use of extended Zernike-Njober theory \cite{Janssen:02}.

Here, we chose to use Gaussian beam mode analysis, an established method based on decomposing fields into Gaussian mode bases (such as the Laguerre-Gauss base), as an efficient way of performing diffraction calculations \cite{Tsigaridas:03}, also for beam propagation of aberrated fields \cite{Trappe:03}. This approach was also recently used for analysis of beam aberrations in confocal Fabry-P\'erot cavities \cite{Mah:19}. Our choice was motivated by the fact that Laguerre-Gaussian beams (\textbf{Figure \ref{fig:AbbIntro}c}) are natural modes for Fabry-P\'erot cavities and, therefore, posses useful characteristics that we will discuss in following sections. Additionally, there is a correspondence between aberration magnitude and LG mode order \cite{Mah:19}. Therefore, small aberrations only cause significant coupling to relatively low order LG modes which allows the analysis to be constrained to a low number of modes. 

We start by defining the general LG mode with indices $l,p$ as:

\begin{equation}
{LG_{lp}}(r,\phi ,z,\lambda)=C_{lp}^{LG}\left({\frac {r{\sqrt {2}}}{w(z)}}\right)^{\!|l|} \\ L_{p}^{|l|}\!\left({\frac {2r^{2}}{w^{2}(z)}}\right)\exp(-il\phi )G(r,z,\lambda)
\end{equation}

where $C_{lp}^{LG}$ is a normalisation constant, $L_p^{|l|}(x)$ is the Laguerre polynomial and $G(r,z,\lambda)$ denotes a general Gaussian beam (see \textbf{Section 1} of the \textbf{Supplementary Information} for full definition):

\begin{equation}
G(r,z,\lambda) = E_{0}\frac {w_{0}}{w(z)}\exp \!\left(\!{\frac {-r^{2}}{w(z)^{2}}}\right) \exp \bigg(-i \bigg(\frac{2\pi}{\lambda}z+{\frac {\pi r^{2}}{\lambda R(z)}}-\psi (z) \bigg)\bigg)
\end{equation}

We now consider an \textit{aberrated} Gaussian beam of the form:

\begin{equation}
G_a(r,\phi,z_0,\lambda)=G(r,z_0,\lambda)\exp \bigg(2\pi i\sum \limits_j \alpha_j Z_j\bigg),
\label{eq:Ga}
\end{equation}

where $\alpha_j$ are the amplitude coefficient of Zernike aberrations expressed in waves and $Z_j$ are Zernike polynomials indexed using the OSA/ANSI standard indices ($Z_j=Z_n^{m}$, where $j=\frac{1}{2}(n(n+2)+m)$). 

\begin{equation}
Z^m_n(r,\phi) = \begin{cases}
      A^m_n R^m_n(r)\cos(m\phi)  & \text{for $m\geqslant0$}\\
      A^m_n R^m_n(r)\sin(m\phi) & \text{for $m<0$},\\
    \end{cases}   
\end{equation}

with

\begin{equation}
R^m_n(r)=\sum \limits_{s=0}^{(n-m)/2} \frac{(-1)^s(n-s)!}{s!((n+m)/2-s)!((n-m)/2-s)!}r^{(n-2s)},
\end{equation}

and $A^m_n$ being a normalisation factor chosen so that:  

\begin{equation}
\max \limits_{r \in [0,1]}Z^m_n(r,\phi)-\min \limits_{r \in [0,1]}Z^m_n(r,\phi)=1.
\end{equation}

This normalisation allows better direct comparison with experimental systems using deformable mirrors or spatial light modulators since their dynamic range is limited by the maximum amplitude of the mode they can display and thus are often calibrated in mode amplitude units. 

We now wish to calculate the coupling or overlap of aberrated Gaussian beam expressed in Zernike modes with the LG modes that propagate inside the FP cavity. For this we seek the electric field $E(r,\phi,z,\lambda)$ of the cavity,

\begin{equation}
E(r,\phi,z,\lambda)= \sum \limits_{l=0}^{\infty} \sum \limits_{p=0}^{\infty} c_{lp}LG_{lp}(r,\phi,z,\lambda),
\label{eq:E}
\end{equation}

where $|c_{lp}|^2$ denotes the fraction of optical power coupled into a particular $LG_{lp}$ mode. These decomposition coefficients can be obtained from:

\begin{equation}
c_{lp}=\iint \limits_{A}{LG_{lp}(r,\phi ,z_0,\lambda)}^*G_a(r,\phi ,z_0,\lambda)dA
\label{eq:clp}
\end{equation}

where A is the aperture over which the field is measured. This now allows us to numerically simulate the F\'abry-Perot interfering field (\textbf{Equation \ref{eq:Efpi}}) by calculating the aberrated electric fields for different reflections in the F\'abry-Perot interferometer.

\section{Effects of beam aberrations on FPI sensitivity}

Our theoretical framework allows us to explore the effects of optical aberrations on the sensitivity of the FP cavity. We start by exploring the properties of an ideal Gaussian beam and note that for a non-aberrated beam all power is confined in the fundamental cavity mode ($LG_{00}$). However, in the presence of aberrations we start to see significant coupling into higher order LG-modes of the cavity (\textbf{Figure \ref{fig:AbbIntro}d}). This effect was also observed in \cite{Mah:19} for confocal cavities, although it does not negatively affect the sensitivity as the ITFs of different LG-modes are fully separated spectrally in this case. 

Next, we consider an ideal FP cavity with flat mirrors and show that for such a cavity illuminated with a perfectly collimated beam aberrations have no effect on sensitivity (\textbf{Appendix A}). In essence, all LG modes show the same transfer function (\textbf{Figure \ref{fig:AbbIntro}e, f}) if the beam diameter is sufficiently large compared to the cavity thickness. This extreme case serves as a check of our analytical result (see \textbf{Appendix A}). However, in realistic experimental conditions \cite{Jathoul:15}, the interrogation light is focused on the FP cavity with spot size $w_0 \leq 50\  \mu m$ which will significantly reduce the Rayleigh range of the beam, effectively leading to spectrally shifted and distorted transfer functions, especially for higher order LG modes (\textbf{Figure \ref{fig:AbbIntro}f}). This in turn, will have serious consequences for the robustness of the cavity as coupling into higher order modes will cause broadening of the ITF and loss of sensitivity.

\begin{figure}[h!]
\includegraphics[width=9.5cm]{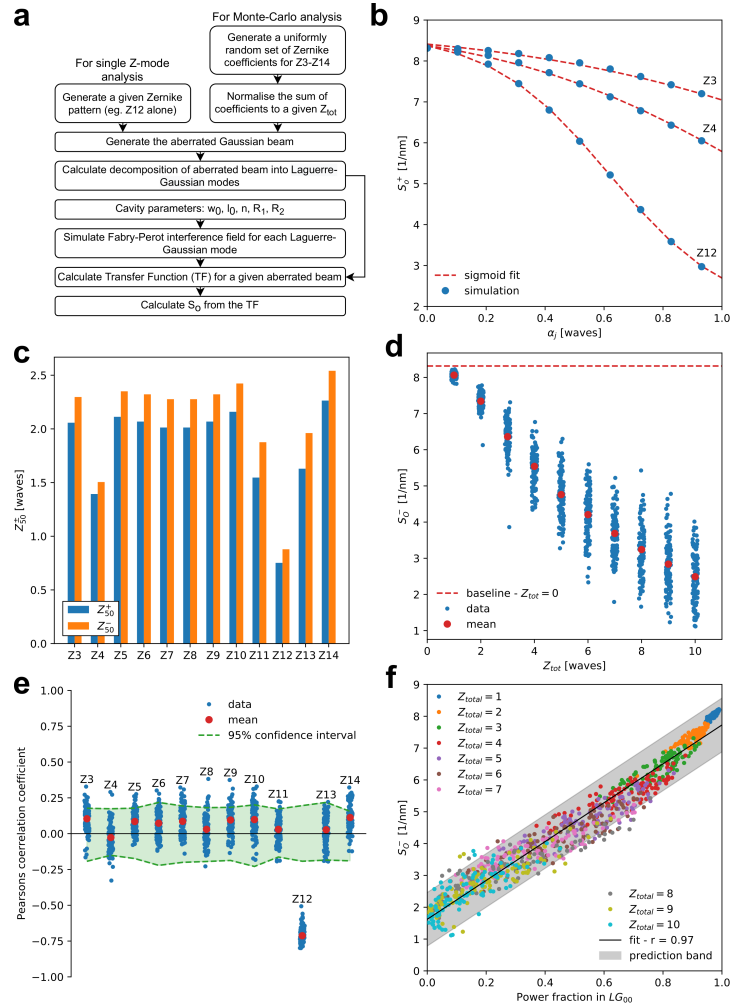}
\centering
\caption{\textbf{a.} Flowchart of the simulation procedure. \textbf{b.} Exemplary graphs of the dependence of $S_o^+$ on the mode amplitude ($\alpha_j$) for modes Z3, Z4 and Z12. All Zernike modes display a sigmoidal dependance. \textbf{c.} Quantification of the effect of different Z-modes on the FPI sensitivity expressed as the aberration magnitude required to lower the sensitivity to 50\% the initial value ($Z_{50}$) for both the falling as well as the rising edge. \textbf{d.} Dependence of the sensitivity ($S_o^+$) on the total aberration ($Z_{tot}$). Each data point is the outcome of an independent simulation with randomly generated aberration. \textbf{e.} Correlation between the magnitude of a particular Z-aberration and the overall sensitivity for a given mixture of aberrations with $Z_{tot}=2$. The slight positive correlation for modes other than Z4 and Z12 is the result of the way we constrain the Zernike amplitudes ($Z_{tot}=const$). Consequently a high amplitude of a weaker mode will reduce the amplitude of modes Z12 and Z4 resulting in a higher overall sensitivity for a constant $Z_{tot}$. \textbf{f.} Correlation between the falling edge sensitivity ($S_o$) and the power conserved in the fundamental $LG_{00}$ mode. The prediction band is the confidence interval for predicting a sensitivity value ($S_o$) given the power contained in the $LG_{00}$ mode, taking into consideration both the confidence of the fit as well as the variance of the data. The band is set for a 95\% confidence interval and thus contains 95\% of the measured points.} 
\label{fig:BeamAb1}
\end{figure}

With this general observation in mind, we now start investigating the effects of \textit{single} Zernike modes on the sensitivity of the FPI. The conceptual procedure of our simulations is given in  (\textbf{Figure \ref{fig:BeamAb1}a}). We observe that while FP sensitivity ($S_o^\pm$) generally declines with increasing amplitude for all aberration modes, the magnitude of their effect is heterogeneous (\textbf{Figure \ref{fig:BeamAb1}b}). To directly compare individual Zernike modes, we calculate the Zernike amplitude ($\alpha_j$) that reduces sensitivity to $50\%$ (\textbf{Figure \ref{fig:BeamAb1}c}). This characteristic point $Z_{50}^\pm$ is important from a practical perspective as it characterises the effective strenght of different modes in degrading the optical sensitivity. Interestingly, we find that Zernike mode 12, i.e. primary spherical aberration (Z12) has the strongest negative effect on sensitivity, followed by Z4 (defocus) and Z11/Z13 (secondary astigmatism). These modes couple most strongly to higher LG modes, presumably because of a combination of factors. Firstly, Zernike polynomials, due to differences in shape, have a variation of volume under the polynomial (VuP) for a constrained maximum amplitude resulting in differences in the overall phase aberration introduced in the beam (\textbf{Figure S1}). Z12 and Z4 show the largest VuP which might explain their strong effect on sensitivity of the FPI. This observation does not, however, fully account for the differences between individual Zernike modes. Additionally, more subtle properties such as the ring-shaped phase of Z12 (\textbf{Figure \ref{fig:AbbIntro}b}) matches well the profiles of higher order LG-modes (\textbf{Figure \ref{fig:AbbIntro}c}), therefore facilitating an efficient coupling. 

In practice, aberrations are never present in isolation, but rather occur as a mixture with varying weights. Hence, we decided to further explore the interactions between different Zernike modes. For this, we pursued a Monte-Carlo approach to analyse these interactions in a high-throughput manner (\textbf{Figure \ref{fig:BeamAb1}a}). In order to constrain optical aberrations in groups we decided to keep the total aberration magnitude constant for each group ($Z_{tot}=\sum \alpha_j=const$).
We observed that the mean sensitivity decreases as the total aberration magnitude increases but also the variance in sensitivity increases for stronger aberrations (\textbf{Figure \ref{fig:BeamAb1}d}). We analysed the source of this variation by calculating the correlation between the sensitivity and aberration magnitude for each of the modes ($\alpha_j$) within a group where $Z_{tot}=const$.
For $Z_{tot}=2$ we observe a strong negative correlation with mode Z12 (Spherical aberration) (\textbf{Figure \ref{fig:BeamAb1}e}) which is in line with the results for single modes in which Z12 has a much stronger impact on the sensitivity than other modes (\textbf{Figure \ref{fig:BeamAb1}c}). 

The most prominent finding of our investigations of beam aberrations is that a single parameter showed a strong linear correlation with the simulated optical sensitivity $S_o^\pm$ (\textbf{Figure \ref{fig:BeamAb1}f}). This parameter is the power fraction conserved in the fundamental $LG_{00}$ mode which seems to suggest that the principal mechanism behind loss of sensitivity is the loss of power in the fundamental mode induced by aberrations.

\section{Effects of cavity aberrations on FPI sensitivity}

\begin{figure}[h!]
\includegraphics[width=10cm]{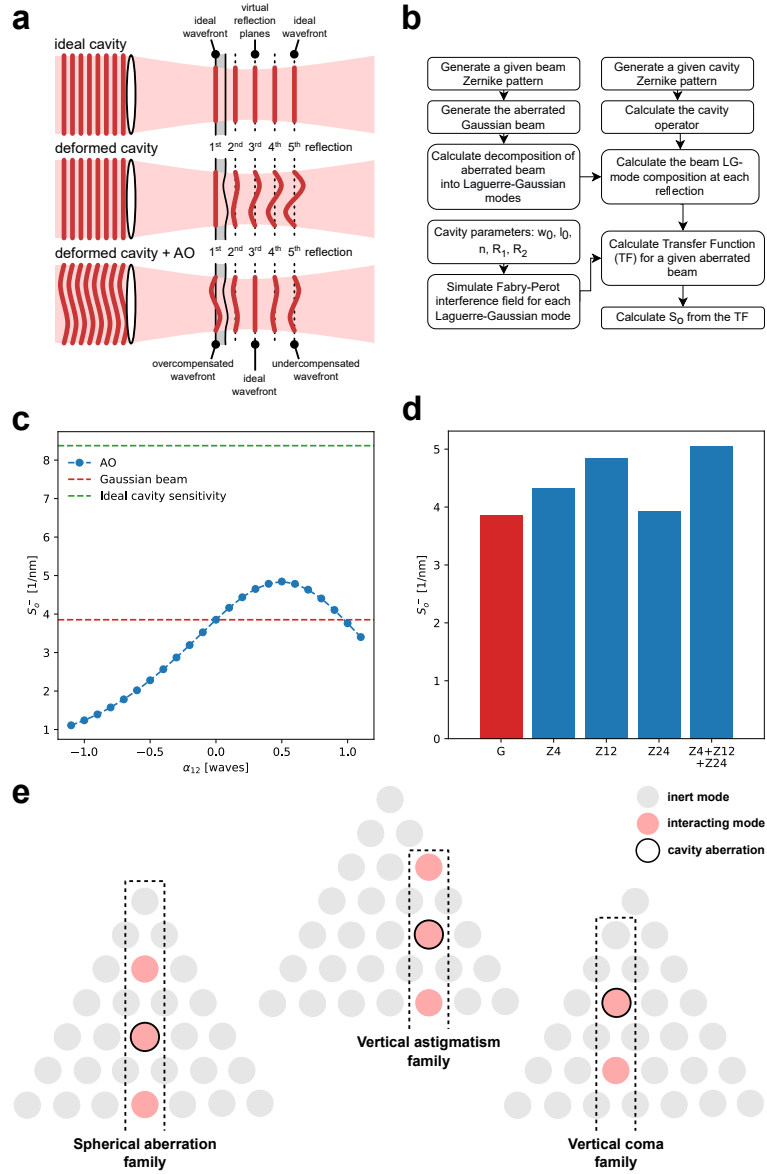}
\centering
\caption{\textbf{a.}  A schematic conceptualising the limitations of adaptive optics in fully compensating the effects of cavity aberrations. \textbf{b.} Flowchart of the simulation procedure. \textbf{c.} AO correction of a $\gamma_{12} = 0.02$ aberrated cavity using mode Z12. \textbf{d.}  Quantification of sensitivity for correcting $\gamma_{12}$ using different Z-modes as well as their combination, which achieves the highest sensitivity. Here the cavity deformation was chosen as $\gamma_{12}=0.02$. \textbf{e.} Z-mode interactions in different deformed cavities. For all cavities: $\gamma = 0.02$.}
\label{fig:AoCav1}
\end{figure}

In the previous section we discussed the effect of beam aberrations on the transfer function of the FPI. However, there is another special class of optical aberrations that needs to be treated separately in our framework, namely, the optical aberrations that are accumulated while the beam is propagating \textit{inside} the cavity. Since the beam makes several round trips ($\sim35-45$ as reported in \cite{Zhang:08}) inside the cavity any phase delay will add up at each reflection. These spatially varying phase delays can e.g. be induced by mirror imperfections (\textbf{Figure \ref{fig:AoCav1}a}) and require the following theoretical treatment.

Here, the main insight is that we can describe the cavity mirror shape again by a combination of Zernike polynomials \cite{Bond:11}:

\begin{equation}
\mathcal{M} (r,\phi,\lambda) = \sum \limits_{j} \gamma_{j}(\lambda) Z_{j}(r,\phi)
\end{equation}

where, $\gamma_j(\lambda)$ is the magnitude of the phase delay introduced to the beam by the mirror deformation expressed in waves. Because this phase delay is introduced at each reflection, it requires a modification of the approach from previous sections. By combining Equation \ref{eq:E} with Equation \ref{eq:Efpi} we can express the interfering electric field $E_{FPI}(r,\phi,\lambda)$ in terms of Laguerre-Gaussian modes:

\begin{equation}
E_{FPI}(r,\phi,\lambda) =\sum \limits_{s=0}^{\infty} c_s \bigg[ {r_1^L}LG_s(r,\phi,0,\lambda)+ \sum \limits_{k=1}^{\infty}\beta_k LG_s(r,\phi,z_k,\lambda') \bigg]
\end{equation}

where $s=(|l|+p)^2+l+|l|+p$ and $l,p$ are LG indices. But, because of cavity aberrations the decomposition of beam into LG modes of weight $c_s$ changes at each reflection $k$. Therefore, the interfering electric field in an aberrated cavity takes the form:

\begin{equation}
E_{FPI}^{\mathcal{M}}(r,\phi,\lambda) =\sum \limits_{s=0}^{\infty} \bigg[ c_s^0{r_1^L}LG_s(r,\phi,0,\lambda)+ \sum \limits_{k=1}^{\infty} c_s^k \beta_k LG_s(r,\phi,z_k,\lambda') \bigg]
\label{eq:E_FPI^M}
\end{equation}

where $|c_s^k|^2$ denotes the fraction of power coupled into mode $LG_s$ at reflection $k$. Because the beam decomposition into LG modes is changing in the cavity at each reflection, the coefficients $c_s^k$ need to be calculated in an iterative fashion:

\begin{equation}
c_{s}^k= \sum \limits_{s'} c_{s'}^{k-1} \iint \limits_{A}{LG_{s}(r,\phi ,z_k,\lambda)} \\ ^*LG_{s'}(r,\phi,z_k,\lambda)\mathcal{M}(r,\phi,\lambda)dA
\end{equation}

This approach is computationally expensive in practice, but can be greatly simplified by assuming that the increase of beam diameter inside the cavity during propagation is small. In our particular case where beam diameter $w_0=50\, \mu m$ and cavity thickness $l_0=20\, \mu m$, the beam will only increase in diameter $\sim 10\, \%$ for an optical path length of 100 reflections. This is negligible for the qualitative conclusions we aim to draw. Because the electric field $E_k$ for reflection $k$ can always be expressed as:

\begin{equation}
E_k (r,\phi,\lambda)=\sum \limits_{s} c_s^k LG_{s}(r,\phi,z_k,\lambda),
\label{eq:Ej_MirrorAb}
\end{equation}

where $s=(|l|+p)^2+l+|l|+p$, we can consider the electric field $E_k$ as a vector in a vector space with LG-modes as an orthonormal base $\{\mathbf{e}_s=LG_{s}(r,\phi,z_k,\lambda)\}$ and $c_s^k$ as coefficients:

\begin{equation}
 \mathbf{E}_{k}=\sum \limits_{s} c_s^k \mathbf{e}_s
\end{equation}

This allows us to define an algebraic operator ($\mathbb{M}$) that describes the mode evolution of the beam inside the cavity (see \textbf{Section 2} of the \textbf{Supplementary Information} for derivation):

\begin{equation}
 \mathbf{E}_{k}=\mathbb{M}\mathbf{E}_{k-1}
\end{equation}

or, in terms of coefficients:

\begin{equation}
c_{s}^{k}=\sum \limits_{s'} \mathbb{M}_{ss'}c_{s'}^{k-1}
\end{equation}

This considerably simplifies the calculation of the amplitude coefficients and power fractions $|c_s^k|^2$ coupled into mode $LG_s$ and the interfering electric field $E_{FPI}^\mathcal{M}(r,\phi,\lambda)$ can then be calculated according to Equation \ref{eq:E_FPI^M}. 

Cavity aberrations have important differences from beam aberrations when considering aberration correction using active wavefront control (\textit{adaptive optics}, AO). Because experimental AO methods only allow wavefront determination and control at a single chosen plane along the optical axis at a time, it is not possible to fully correct cavity aberrations as they evolve through each reflection (\textbf{Figure \ref{fig:AoCav1}a}). Thus, we aim to explore to which extent cavity aberrations can be corrected. In particular, we perform \textit{in silico} AO experiments for various cavity deformations by establishing an appropriate simulation pipeline (\textbf{Figure  \ref{fig:AoCav1}b}). Utilizing our simulation routine, we first investigated and characterized the capability of individual Z-modes to correct a given cavity deformation (\textbf{Figure  \ref{fig:AoCav1}c}, \textbf{Figure S2}).

Here, we found that more than a single Z-mode can interact with the cavity, even when the cavity is deformed using only a single Zernike polynomial. This led to the observation that eg. $\gamma_{12}$ can be corrected by $Z4$, $Z12$ as well as $Z24$ (\textbf{Figure \ref{fig:AoCav1}d}, \textbf{Figure S2}). We explored this in a more rigorous fashion and found that Z-modes within the same 'family' have the ability to partially compensate for each other (\textbf{Figure \ref{fig:AoCav1}e}). Furthermore, when combined together they can even further improve FP sensitivity (\textbf{Figure \ref{fig:AoCav1}d}).

\section{Active and passive aberration correction in FPI systems}
\subsection*{Beam aberrations}
After discussing the effects of both beam and cavity aberrations in the previous sections, we now proceed to explore potential approaches for correcting them, and thus recovering the loss of sensitivity. Typically, beam aberrations in microscopy setups are addressed with the use of deformable mirrors (DMs) or spatial light modulators (SLMs) which can apply spatially varying, controlled phase delays.
However, based on our simulations we also hypothesized that a much simpler approach could be effective. Since the loss of sensitivity is caused mainly by leakage of power to higher-order cavity modes (\textbf{Figure \ref{fig:BeamAb1}f}), we speculated that sensitivity could be improved by mode filtering, e.g. by using a passive element such as a single mode fibre. This serves to reject all the reflected light that propagates outside the fundamental $LG_{00}$ (i.e. Gaussian) mode from reaching the detector. 

We tested our hypothesis both in simulations as well as experimentally. For our simulations we calculated the coupling of light power from the interfering field $E_{FPI}$ into the fundamental Gaussian mode:

\begin{equation}
I_{FPI}^{SM}(\lambda)=\bigg\rvert \iint \limits_{A} E_{FPI}(r,\phi,\lambda)^*G(r,\phi,0,\lambda)dA \ \bigg\rvert^2
\label{eq:Ifpi}
\end{equation}

We observed that this mode filtering approach has the same effect as active AO correction in recovering the ideal transfer function (\textbf{Figure \ref{fig:BeamAb2}a}), while being much simpler to implement experimentally. The only disadvantage is that for large aberration the power loss through filtering becomes significant (\textbf{Figure \ref{fig:BeamAb2}b}). 

\subsection*{Cavity aberrations}
In the previous section we treated the effects of beam aberrations and described potential ways to tackle them using both active and passive aberration correction. In this section we will discuss the case of cavity aberrations and how both active and passive correction could synergise. 

We start by stating that similarly to active correction, passive mode filtering will also not be fully effective in tackling sensitivity loss due to cavity aberrations, since the power distribution between the modes will change in-between reflections, thus distorting the Gaussian mode interference pattern (\textbf{Figure  \ref{fig:BeamAb2}c}). This highlights the importance of manufacturing cavities with high uniformity of thickness as cavity deformations cannot be fully corrected and thus lead to an irreversible loss of sensitivity, unless techniques to actively alter the cavity structure locally are implemented. As some of such techniques are being actively developed \cite{Villringer:20, Chen:20} it remains to be seen if they prove to be effective in tackling cavity induced loss of sensitivity.

Our simulations show that both AO and single-mode filtering (SM) individually do improve sensitivity over the aberrated case (\textbf{Figure  \ref{fig:BeamAb2}d,e}). Contrary to beam aberrations, for cavity aberrations, AO and SM have different correction mechanisms which can in fact complement each other. Interestingly if AO is performed \textit{while} the beam is being mode filtered we can achieve even higher sensitivity than by mode filtering the AO corrected beam ($AO(SM)>SM(AO)$). The reason for this lies in the fact that for $AO(SM)$ only the fundamental mode is optimized for amplitude and phase distribution between reflections. This yields a different overall correction compared to optimizing reflections for all LG modes in case of $AO$. However, similar to the case of beam aberrations, passive aberration correction comes at the price of rejecting significant amounts of light for large aberrations (\textbf{Figure  \ref{fig:BeamAb2}f}).

\subsection*{Experimental validation}

To validate our theoretical findings we also performed experiments using an all-optical photoacoustic tomography setup based on a Fabry-P\'erot sensor that conceptually follows  \cite{Zhang:08} but was modified by adding an adaptive optics module (\textbf{Figure S3a}) consisting of a deformable mirror conjugated to the back focal plane of the scan lens. Mode filtering was realised experimentally by coupling the reflected light into a single-mode fibre (for more details see \textbf{Section 3} of the \textbf{Supplementary Information}). Optical sensitivity is calculated by fitting a numerical FP transfer function\cite{Varu:14} to the raw data (\textbf{Equation \ref{eq:SO}, Figure S3b}). 

\begin{figure}
\includegraphics[width=11cm]{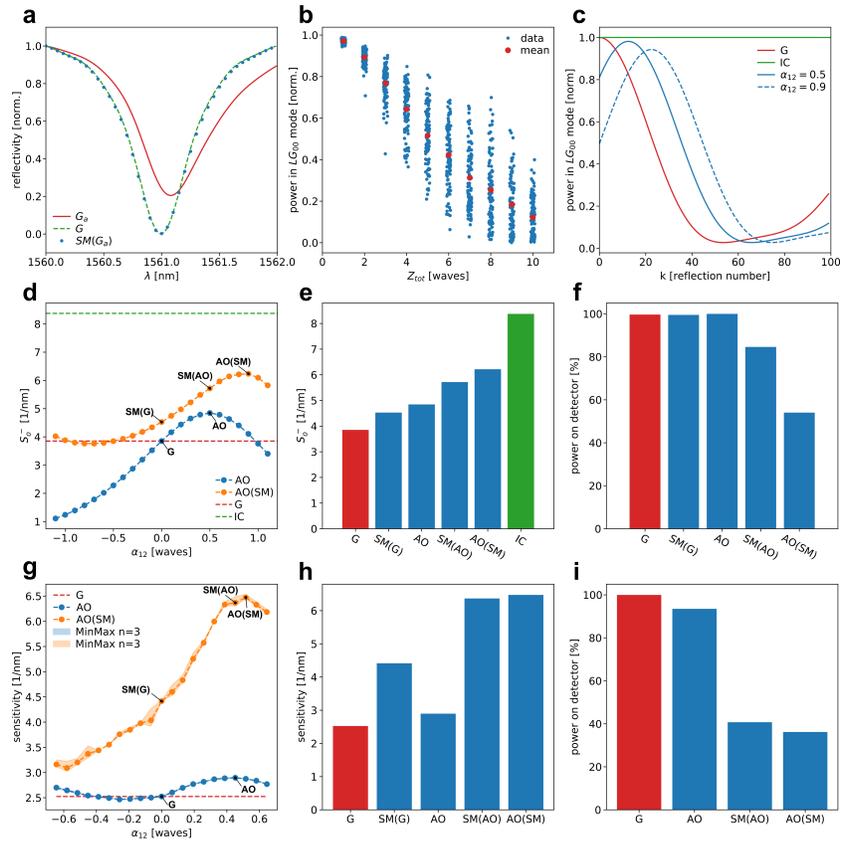}
\centering
\caption{Experimental characterization of FPI sensitivity improvement using active and passive aberration correction. \textbf{a.} Exemplary plot showing the simulated effects of mode filtering on the transfer function of the FPI. \textbf{G\textsubscript{a}} - Aberrated Gaussian beam, \textbf{G} - ideal Gaussian beam, \textbf{SM(G\textsubscript{a})} - mode filtered aberrated Gaussian beam. \textbf{b.} Dependence of the power contained in the fundamental mode ($LG_{00}$) on the total aberration ($Z_{tot}$). For large aberrations passive AO rejects significant portions of the light. \textbf{c.} Dependence of power contained in the fundamental $LG_{00}$ mode on the reflection number $k$ inside the FPI. \textbf{d.} Simulation of AO correction using mode Z12 comparing the correction with and without mode filtering. Cavity parameters: $w_0=50 \mu m$, $l_0=20 \mu m$, $R_{1/2}=0.95$, $\gamma_{12}=0.2$. \textbf{e.} Comparison of the 5 characteristic points from \textbf{d}, showing that AO on fiber-coupled detection achieves highest sensitivity. \textbf{f.} Comparison of theoretical power on the detector for different AO methods. Combination of AO and SM leads to considerable power loss. \textbf{g.} Experimental AO correction using mode Z12 with and without additional mode filtering. Cavity parameters: $w_0 \sim 50 \mu m$, $l_0 \sim 22 \mu m$, $R_{1/2} \sim 0.98$, $\gamma - $unknown. \textbf{h.} Comparison of the 5 characteristic points from \textbf{g}. Combinations of AO and SM can increase sensitivity almost three-fold. \textbf{i.} Comparison of experimentally measured power on the detector for different AO methods. $SM(AO)$ and $AO(SM)$ are normalised to the fibre coupled power at $\alpha_{12}=0.0$ to disentangle the experimental power loss due to fibre coupling. \textbf{G} - Gaussian beam, \textbf{IC} - ideal cavity, \textbf{AO} - AO corrected beam, \textbf{SM(G)} - mode filtered Gaussian beam, \textbf{SM(AO)} - mode filtered AO corrected beam, \textbf{AO(SM)} - AO correction performed while mode filtering.}
\label{fig:BeamAb2}
\end{figure}

It is important to note here that in our experiments both the exact beam ($\alpha_j$) as well as cavity ($\gamma_j$) aberrations remain in principle unknown, therefore preventing precise modeling of the experimental situation. Nevertheless, qualitative comparisons can be made to gain intuitive insights into the system. In our experiments, we observed that mode filtering indeed increases the sensitivity compared to performing adaptive optics only which stands in agreement with our simulations (\textbf{Figure  \ref{fig:BeamAb2}g}). Furthermore, the quantification of the characteristic points also shows similarities with the relative sensitivity improvements mostly conserved $AO(SM)>SM(AO)>AO/SM(G)>G$ (\textbf{Figure  \ref{fig:BeamAb2}h}). Differently to the simulation $AO < SM(G)$, however, this may be due to contributions from other cavity and beam aberrations which are not experimentally characterised, as mentioned above. Finally, we quantified the power loss for various AO approaches, which also shows qualitative agreements with our simulations (i.e. $AO(SM)<SM(AO)<G$ , see \textbf{Figure  \ref{fig:BeamAb2}i}).

\section{Discussion}
 We have shown theoretically as well as experimentally that active and passive techniques to correct aberrations have the capability to increase optical sensitivity of FP-based sensors. One unexpected finding of our investigation is the fact that passive mode filtering can achieve significant gains in optical sensitivity. We expect this to have impact in practical realizations of FPIs, such as in photoacoustic imaging, because of the simplicity and ease of its experimental implementation. The only disadvantage and limitation lies in the fact that, depending on the amplitude of the aberration, a considerable part of the interrogating laser light might be rejected. We further found that, in more realistic cases when both cavity and beam aberrations are present, combining active and passive techniques yields the overall best improvements in sensitivity (\textbf{Figure \ref{fig:BeamAb2}e,h}). As again power loss of the passive filtering might be limiting, the optimal solution may thus not only depend on the effective increase in sensitivity, but also on the effects due to reduced power and signal-to-noise ratio on the detector side. This might therefore require a more complex optimisation metric which takes into account these additional considerations concerning signal and detector noise sources. 
 
 An important insight of our work with much broader applicability and potential impact is our theoretical observation that higher order aberrations can actually be partially corrected by lower order modes (eg. $Z4$ can correct $\gamma_{12}$). This has important practical implications since it might allow to use active optical elements such as deformable mirror with lesser degrees of freedom. This would greatly reduce both the cost as well as technical complexity of experimental AO correction, and thus might lead to a more widely uptake in the field. Furthermore, we note that our theoretical approach can also be used to investigate other metrics of the FP transfer function, such as visibility or linewidth, which are important for other FP-based sensing applications.
 
 Moreover, recent work has shown that the use of non-Gaussian beams can in principle further increase FP measurement sensitivity, e.g. by utilizing $LG_{33}$ modes in LIGO detectors \cite{Noack:17}, or Bessel beams in FP based pressure sensing \cite{Sheppard:20}. 
 Our model could therefore be used to further evaluate and explore the robustness of these and other non-Gaussian beams against beam and cavity aberrations in FPI interrogation. Finally, we expect that our theoretical framework will find application beyond photoacoustics-based pressure sensing. For example, our general finding might be also further explored for other imaging modalities such as multi-photon microscopy where AO is utilized to correct more complex aberrations induced by living tissue \cite{Ji:17}, or to increase the spectral resolution in VIPA-based, non-confocal Brillouin spectrometers\cite{Zhang:16}. 

 See Supplementary document for supporting content.
 
\section{Appendix A. Limit case for ideal FP cavity illuminated with a non-diverging beam}

The properties of a planar Fabry-P\'erot cavity illuminated with non-diverging beams can be analysed analytically to evaluate the effect of optical aberrations on the ITF. We aim to calculate the transfer function $I_{FPI}^{ND}(\lambda)$ where we use the superscript $ND$ to denote non-divergence of the beam. From Equation \ref{eq:I_FPI} we know that:

\begin{equation}
I_{FPI}^{ND} (\lambda) = \iint \limits_{A} E_{FPI}^{ND}(r,\phi,\lambda)^*E_{FPI}^{ND}(r,\phi,\lambda)  \,dA
\label{eq:I_FPI^ND}
\end{equation}

And by combining Equation \ref{eq:E} with Equation \ref{eq:Efpi} we can express the interfering electric field $E_{FPI}^{ND}$ in terms of Laguerre-Gaussian modes:

\begin{equation}
E_{FPI}^{ND}(r,\phi,\lambda) =\sum \limits_{l=0}^{\infty} \sum \limits_{p=0}^{\infty} c_{lp} \sum \limits_{i=0}^{\infty}\beta_i LG_{lp}^{ND}(r,\phi,z_i,\lambda)
\label{eq:E_FPI^ND}
\end{equation}

where,

\begin{equation}
\beta_i = \begin{cases}
      r_1^L & \text{for $i=0$}\\
      2{(t_1)}^2{(r_1^R)}^{i-1}{(r_2^L)}^{i} & \text{for $i>0$}\\
    \end{cases}   
\end{equation}

Now we need to explore the properties of of non-diverging LG modes. We start by defining a non-diverging LG mode as the limit when the Rayleigh range of the beam ($z_{\mathrm{R}}$) approaches infinity:

\begin{multline}
LG_{lp}^{ND}(r,\phi ,z,\lambda) = \lim_{z_{\mathrm{R}} \rightarrow{} \infty} {LG_{lp}}(r,\phi ,z,\lambda)= \lim_{z_{\mathrm{R}} \rightarrow{} \infty} C_{lp}^{LG}\left({\frac {r{\sqrt {2}}}{w(z)}}\right)^{|l|} L_{p}^{|l|} \bigg({\frac {2r^{2}}{w^{2}(z)}} \bigg)
\\ \exp(-il\phi )\frac {w_{0}}{w(z)}\exp \bigg({\frac {-r^{2}}{w(z)^{2}}} \bigg) \exp \bigg(-i \bigg(\frac{2\pi}{\lambda}z+{\frac {\pi r^{2}}{\lambda R(z)}}-\psi (z) \bigg)\bigg)
\label{eq:LGC}
\end{multline}

We will now explore the limits of all the parts dependent on $z_{\mathrm{R}}$ separately:

\begin{equation}
\lim_{z_{\mathrm{R}} \rightarrow{} \infty} w(z)= \lim_{z_{\mathrm{R}} \rightarrow{} \infty} w_{0}\sqrt {1+{\left({\frac{z}{z_{\mathrm{R}}}}\right)}^{2}} = w_{0}
\end{equation}

\begin{equation}
\lim_{z_{\mathrm{R}} \rightarrow{} \infty} \psi(z) = \lim_{z_{\mathrm{R}} \rightarrow{} \infty} \arctan \bigg(\frac{z}{z_{\mathrm{R}}} \bigg) = 0
\end{equation}

The value of $r$ is also indirectly dependent on $z_{\mathrm{R}}$ because $r \sim w_0$ for proper beam sampling and from \textbf{Equation S5} we know that $w_0^2 \sim z_{\mathrm{R}}$ so $r^2 \sim z_{\mathrm{R}} $:

\begin{equation}
\lim_{z_{\mathrm{R}} \rightarrow{} \infty} \frac{r^2}{ R_z(z)}=\lim_{z_{\mathrm{R}} \rightarrow{} \infty} \frac{z{z}_{\mathrm{R}}}{z^2+z_{\mathrm{R}}^2} = 0
\end{equation}

With these we come back to Equation \ref{eq:LGC}:
\begin{multline}
LG_{lp}^{ND}(r,\phi ,z,\lambda) = \lim_{z_{\mathrm{R}} \rightarrow{} \infty} {LG_{lp}}(r,\phi ,z,\lambda)= \lim_{z_{\mathrm{R}} \rightarrow{} \infty} C_{lp}^{LG}\left({\frac {r{\sqrt {2}}}{w(z)}}\right)^{|l|} L_{p}^{|l|} \bigg({\frac {2r^{2}}{w^{2}(z)}} \bigg)
\\ \exp(-il\phi )\frac {w_{0}}{w(z)}\exp \bigg({\frac {-r^{2}}{w(z)^{2}}} \bigg) \exp \bigg(-i \bigg(\frac{2\pi}{\lambda}z+{\frac {\pi r^{2}}{\lambda R(z)}}-\psi (z) \bigg)\bigg)=
\\ C_{lp}^{LG}\left({\frac {r{\sqrt {2}}}{w_0}}\right)^{|l|} L_{p}^{|l|} \bigg({\frac {2r^{2}}{{w_0}^{2}}} \bigg) \exp(-il\phi ) \exp \bigg({\frac {-r^{2}}{{w_0}^{2}}} \bigg) \exp \bigg(-i\frac{2\pi}{\lambda}z\bigg)
\label{eq:LGC2}
\end{multline}

where we reach our first conclusion by observing that $LG_{lp}^{ND}(r,\phi ,z,\lambda)$ is separable:

\begin{equation}
    LG_{lp}^{ND}(r,\phi ,z,\lambda)=LG_{lp}^{ND}(r)LG_{lp}^{ND}(\phi)LG_{lp}^{ND}(z,\lambda)
\end{equation}

which leads to the first property of non-diverging LG modes:

\begin{equation}
LG_{lp}^{ND}(r,\phi ,z_i,\lambda) = LG_{lp}^{ND}(r,\phi ,z_{i'},\lambda) \exp \bigg(- \frac{2\pi i (z_i-z_{i'})}{\lambda} \bigg),
\end{equation}

The second property flows directly from orthonormality of LG-modes:

\begin{equation}
\delta_{l'l}\delta_{p'p}=\iint \limits_{A}{LG_{lp}^{ND}(r,\phi ,z_i,\lambda)} \\ {LG_{l'p'}^{ND}(r,\phi ,z_i,\lambda)^*}dA.
\end{equation}

We combine these to achieve:

\begin{equation}
\delta_{l'l}\delta_{p'p}exp \bigg(- \frac{2\pi i (z_i-z_{i'})}{\lambda} \bigg)=\iint \limits_{A}{LG_{lp}^{ND}(r,\phi ,z_{i},\lambda)} \\ {LG_{l'p'}^{ND}(r,\phi ,z_{i'},\lambda)^*}dA
\label{eq:LG_property}
\end{equation}

Now we return to Equation \ref{eq:I_FPI^ND} and proceed to calculate the transfer function of an ideal FP cavity:

\begin{multline}
I_{FPI}^{ND}(\lambda) = \iint \limits_{A} E_{FPI}^{ND}(r,\phi,\lambda)^*E_{FPI}^{ND}(r,\phi,\lambda) \,dA \overset{(\ref{eq:E_FPI^ND})}{=} 
\\ \iint \limits_{A} \bigg( \sum \limits_{l=0}^{\infty} \sum \limits_{p=0}^{\infty} c_{lp} \sum \limits_{j=0}^{\infty}\beta_j LG_{lp}^{ND}(r,\phi,z_j,\lambda) \bigg) \bigg( \sum \limits_{l'=0}^{\infty} \sum \limits_{p'=0}^{\infty} c_{l'p'} \sum \limits_{j'=0}^{\infty}\beta_{j'} LG_{l'p'}^{ND}(r,\phi,z_{j'},\lambda) \bigg)^* dA =
\\ \sum \limits_{l=0}^{\infty} \sum \limits_{p=0}^{\infty} \sum \limits_{l'=0}^{\infty} \sum \limits_{p'=0}^{\infty} c_{l'p'}^* c_{lp} \sum \limits_{j=0}^{\infty} \sum \limits_{j'=0}^{\infty}\beta_{j'}^* \beta_j \iint \limits_{A} LG_{lp}^{ND}(r,\phi,z_j,\lambda) LG_{l'p'}^{ND}(r,\phi,z_{j'},\lambda)^* dA  \overset{(\ref{eq:LG_property})}{=} 
\\ \sum \limits_{l=0}^{\infty} \sum \limits_{p=0}^{\infty} \sum \limits_{l'=0}^{\infty} \sum \limits_{p'=0}^{\infty} c_{l'p'}^* c_{lp} \sum \limits_{j=0}^{\infty} \sum \limits_{j'=0}^{\infty}\beta_{j'}^* \beta_j \delta_{l'l}\delta_{p'p} \exp \bigg(- \frac{2\pi i (z_j-z_{j'})}{\lambda} \bigg) =
\\ \sum \limits_{l=0}^{\infty} \sum \limits_{p=0}^{\infty} |c_{lp}|^2 \sum \limits_{j=0}^{\infty} \sum \limits_{j'=0}^{\infty}\beta_{j'}^* \beta_j \exp \bigg(- \frac{2\pi i (z_j-z_{j'})}{\lambda} \bigg)=
\\ \sum \limits_{l=0}^{\infty} \sum \limits_{p=0}^{\infty} |c_{lp}|^2 \sum \limits_{j=0}^{\infty}\beta_j \exp \bigg(- \frac{2\pi i z_j}{\lambda} \bigg) \sum \limits_{j'=0}^{\infty}\beta_{j'}^* \exp \bigg(- \frac{2\pi i z_{j'}}{\lambda} \bigg)^*
\end{multline}

Taking in consideration the following:

\begin{equation}
\sum \limits_{j'=0}^{\infty}\beta_{j} \exp \bigg(- \frac{2\pi i z_{j}}{\lambda} \bigg) = E_{FPI}^{Airy}(\lambda)
\end{equation}

where $E_{FPI}^{Airy}(\lambda)$ is the well known solution for an ideal FPI illuminated with a plane wave. And normalisation of the power of the beam:

\begin{equation}
\sum \limits_{l=0}^{\infty} \sum \limits_{p=0}^{\infty} |c_{lp}|^2 = 1
\end{equation}

We conclude:

\begin{equation}
I_{FPI}^{ND}(\lambda)=E_{FPI}^{Airy}(\lambda)E_{FPI}^{Airy}(\lambda)^*=I_{FPI}^{Airy}(\lambda)
\end{equation}

This results shows that if the beam is non-diverging it will create an ideal Airy interference pattern inside a Fabry-P\'erot cavity regardless of it's decomposition into LG-modes. As any beam can be represented as a linear combination of LG-modes this shows that an Fabry-P\'erot cavity illuminated with a non-diverging beam is inherently resistant to beam aberrations. Importantly, however, this conclusion does not hold for confocal cavities because interference patterns of different LG-modes experience a spectral shift due to the curvature of the mirrors.

\section*{Funding}
This work was supported by the European Molecular Biology Laboratory (EMBL).

\section*{Acknowledgments}
We acknowledge Florian Mathies, Johannes Zimmermann and Gerardo Hernandez-Sosa from InnovationLab Heidelberg as well as Karl-Phillip Strunk and Jana Zaumseil from Centre for Advanced Materials, Heidelberg University for help with manufacturing of the Fabry-P\'erot interferometers with elastic cavities used in this work. We acknowledge the Mechanical Workshop at EMBL Heidelberg for manufacturing custom opto-mechanical components required for the experimental setup.

\section*{Disclosures}

The authors declare no conflicts of interest.

\bibliography{sample}

\newpage
\clearpage

\section{Supplementary material}
\beginsupplement

This document provides supplementary information on the main manuscript "Effects of optical aberrations on sensitivity of planar Fabry-P\'erot cavities". It contains full definitions of functions and operators used in our theoretical approach. Furthermore, it includes more details on the effect of various Zernike aberrations on FPI sensitivity and describes the optical setup used to acquire the experimental data presented in the main manuscript.

\section{Supplementary Definitions}

The definition of a general Gaussian beam as used in our theoretical approach (Section 3, main manuscript):

\begin{equation}
G(r,z,\lambda) = E_{0}\frac {w_{0}}{w(z)}\exp \!\left(\!{\frac {-r^{2}}{w(z)^{2}}}\right) \exp \bigg(-i \bigg(\frac{2\pi}{\lambda}z+{\frac {\pi r^{2}}{\lambda R(z)}}-\psi (z) \bigg)\bigg)
\end{equation}

with $w(z)$ the local beam radius,

\begin{equation}
w(z)=w_0\sqrt{1+\bigg( \frac{z}{z_{\mathrm{R}}} \bigg)^2},
\end{equation}

$R(z)$ the local beam curvature,

\begin{equation}
R(z)=z\bigg[1+\bigg(\frac{z_{\mathrm{R}}}{z} \bigg)^2\bigg],
\end{equation}

$\psi(z)$ the Gouy phase
\begin{equation}
\psi(z)=\arctan \bigg(\frac{z}{z_{\mathrm{R}}} \bigg),
\end{equation}

and $z_{\mathrm{R}}$ the Rayleigh range of the beam
\begin{equation}
z_{\mathrm{R}}=\frac{\pi {w_0}^2 n_0}{\lambda}.
\label{eq:Zr}
\end{equation}

Here, $n_0$ is the refractive index of the propagation medium and $w_0$ is the beam radius in focus. 

\section{Cavity evolution operator}

We can define the mode evolution operator $\mathbb{M}$ by (Section 5, main manuscript):

\begin{equation}
c_{s}^{k}=\sum \limits_{s'} \mathbb{M}_{ss'}c_{s'}^{k-1}
\label{eq:OperatorDefinition}
\end{equation}

We calculate the operator elements $\mathbb{M}_{ss'}$ by determining the cross-coupling between all LG modes and their aberrated counterparts:

\begin{equation}
\mathbb{M}_{ss'}=\iint \limits_{A}{LG_{s}(r,\phi ,z_0,\lambda)} \\ ^*LG_{s'}^{\mathcal{M}}(r,\phi,z_0,\lambda))dA
\end{equation}

where $LG_{s}^{\mathcal{M}}(r,\phi,z_0,\lambda)$ is the mirror aberrated LG mode:

\begin{equation}
LG_{s}^{\mathcal{M}}(r,\phi,z_0,\lambda)=LG_{s}(r,\phi,z_0,\lambda) \\ \exp (2\pi i\mathcal{M}  (r,\phi,\lambda)).
\end{equation}

One can then use Equation 17 to calculate the fields required for Equation 4. The initial mode decomposition of the beam $c_{s}^0$ needs to be calculated from the input field: 

\begin{equation}
c_{s}^0=\iint \limits_{A}{LG_{s}(r,\phi ,z_0,\lambda)}^*G_a(r,\phi,z_0,\lambda)) \\ \exp (2\pi i\mathcal{M}  (r,\phi))dA
\label{eq:Clp0}
\end{equation}

where $G_a$ can be an aberrated beam from Equation 7 or an non-aberrated beam by setting $\sum |\alpha_j|=0$.

\section{Supplementary Experimental Methods}

\begin{figure}[h!]
\includegraphics[width=10cm]{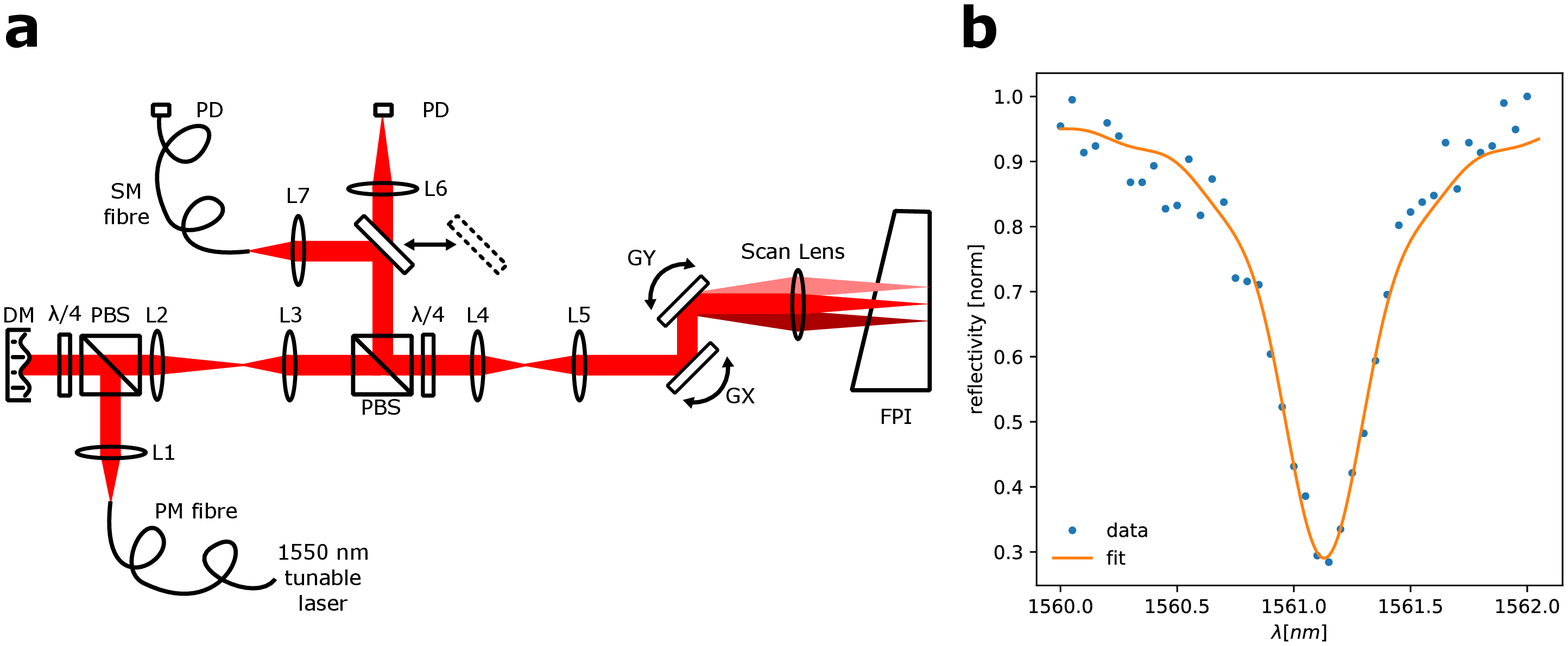}
\centering
\caption{\textbf{a.} Schematics of the experimental setup. \textbf{L1} - collimates the output of the interrogation laser's fiber to match the diameter of the active aperture of the deformable mirror (\textbf{DM}). Two optical relays (\textbf{L2-L3} and \textbf{L4-L5}) then reduce the beam diameter appropriately to achieve $\sim 50 \mu m$ spot radius on the Fabry-P\'erot Interferometer (\textbf{FPI}). \textbf{GX, GY} - galvanometric mirrors, \textbf{PD} - photodiode, \textbf{PBS} - polarising beamsplitter, $\boldsymbol{\lambda} \mathbf{/4}$ - quarter-wave plate,  \textbf{PM fibre} - polarisation-maintaining fibre,  \textbf{SM fibre} - single-mode fibre.  \textbf{b.} Exemplary data of an experimental, mode filtered ITF, showing good agreement between experimental data points and respective fit from which the optical sensitivity is inferred.}
\label{fig:Supp_Setup}
\end{figure}

The optical setup used for the experiments presented in this work is outlined in \textbf{Figure \ref{fig:Supp_Setup}a}. First, the output of the interrogation laser is collimated and its size matched to the diameter of the active aperture ($\sim 10$ mm) of the deformable mirror (\textbf{DM}, DMP40/M-P01, Thorlabs). Two relays (\textbf{L2-L3} and \textbf{L4-L5}) then reduce the beam diameter by 0.6x and 0.625x, respectively, to match the required NA for the scan lens (TSL-1550-15-80, Wavelength Opto-Electronic) to achieve a $\sim 50 \mu m$ spot radius on the Fabry-P\'erot Interferometer (\textbf{FPI}). The back reflected light is redirected by a quarter-waveplate ($\boldsymbol{\lambda} \mathbf{/4}$) and polarising beamsplitter (\textbf{PBS}) to the detection path and then either directly focused, or fiber coupled into a single-mode fibre, before detection by a photodiode (\textbf{PD}, PDA05CF2, Thorlabs). The DM was factory-precalibrated to display Zernike modes 3 to 15. The tuneable interrogation laser, DM and data acquisition (NI-6259, National Instruments) are controlled by a custom written LabView software. The FP interferometer transfer function (ITF - \textbf{Figure \ref{fig:Supp_Setup}b}) is acquired by first setting a particular Zernike mode pattern on the DM and then tuning the wavelength of the laser in a stepwise manner to avoid spectra deformations connected to continuous wavelength sweeping.

For accurately estimate the FP sensitivity, the ITF data is fitted numerically considering the function of a Gaussian beam propagating in an ideal cavity and varying the reflectivity of the two mirrors ($R_1$,$R_2$ - \textbf{Figure \ref{fig:Supp_Setup}b}). This fitting approach \cite{Varu:14} showed good performance and can be computed efficiently.

\newpage
\section{Supplementary Figures}

\begin{figure}[h!]
\includegraphics[width=8cm]{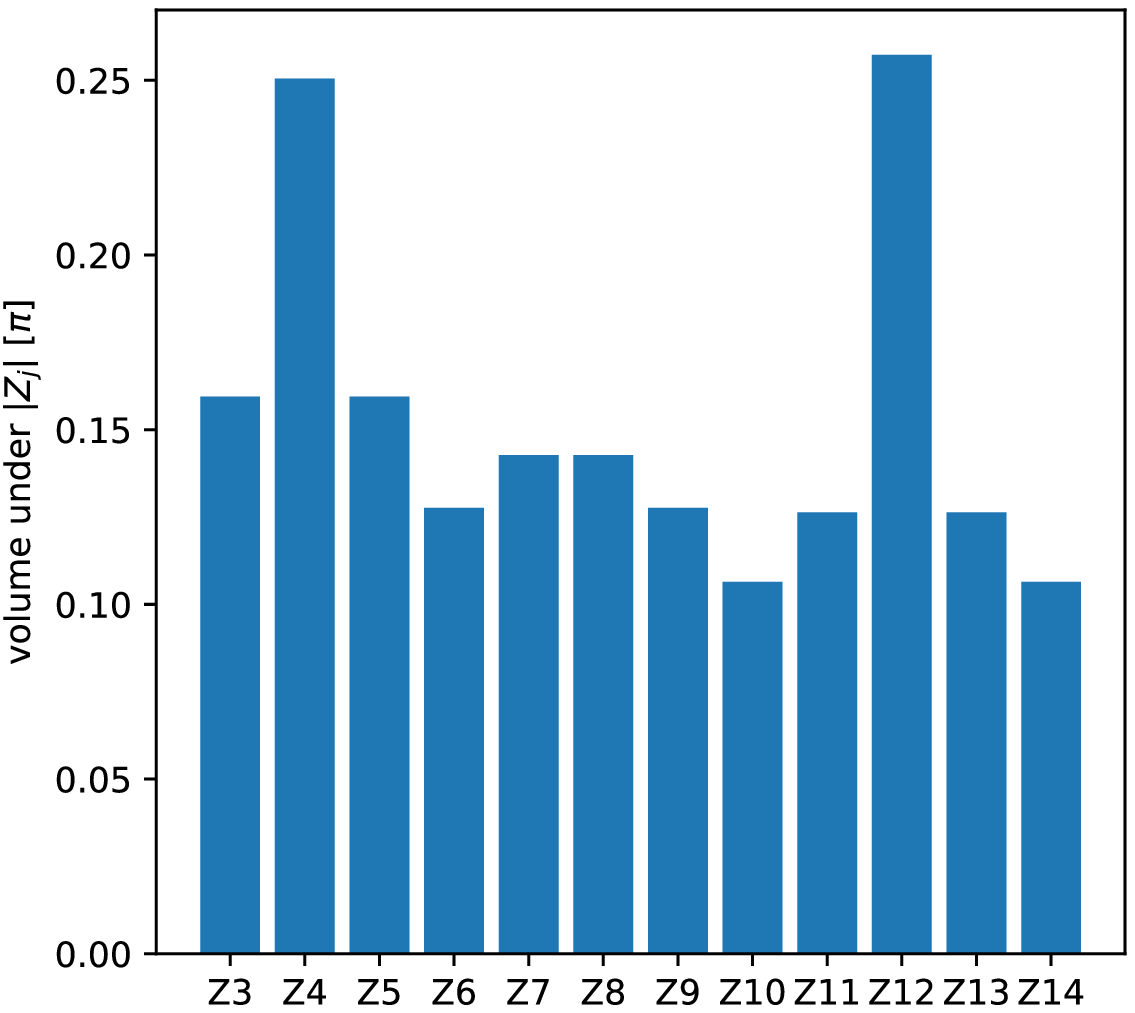}
\centering
\caption{Volume under the absolute value of the Zernike polynomial (VuP) showing differences which may contribute to differential effects of Zernike aberrations on FPI sensitivity ($VuP_j=\int_0^1 \int_{-\pi}^{\pi} |Z_j(r,\phi)|rd\phi dr$).}
\end{figure}

\begin{figure}[h!]
\includegraphics[width=10cm]{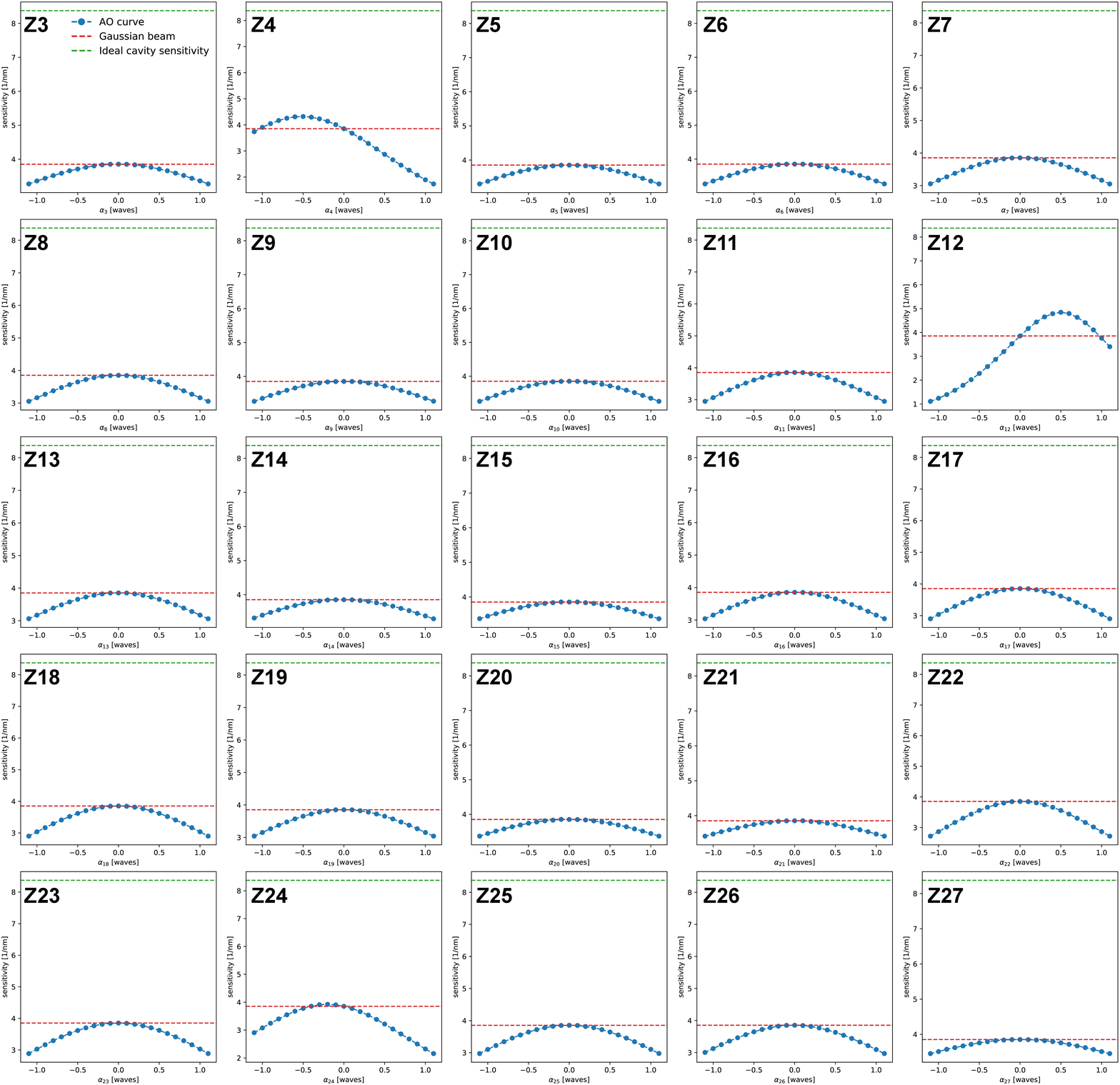}
\centering
\caption{Dependence of optical sensitivity ($S_o^-$) on the amplitude of individual Zernike aberrations ($\alpha_j$) for the case of a $\gamma_{12}=0.02$ aberrated cavity.}
\end{figure}

\end{document}